\documentclass[preprint]{aastex}

\def\lesssim{\mathrel{\hbox{\rlap{\hbox{\lower4pt\hbox{$\sim$}}}\hbox{$<$}}}}
\def\gtrsim{\mathrel{\hbox{\rlap{\hbox{\lower4pt\hbox{$\sim$}}}\hbox{$>$}}}}

\usepackage{graphicx}
\usepackage{color}
\usepackage{natbib}
\begin{document}

\title{RADIATIVE EFFICIENCY AND THERMAL SPECTRUM OF ACCRETION ONTO SCHWARZSCHILD BLACK HOLES}

\author{Scott C. Noble}
\affil{Center for Computational Relativity and Gravitation\\
Rochester Institute of Technology\\
Rochester, NY 14623}
\email{scn@astro.rit.edu}

\and

\author{Julian H. Krolik}
\affil{Physics and Astronomy Department\\
Johns Hopkins University\\ 
Baltimore, MD 21218}
\email{jhk@jhu.edu}

\and

\author{Jeremy D. Schnittman}
\affil{NASA/Goddard Space Flight Center\\
Greenbelt, MD 20771}
\email{jeremy.d.schnittman@nasa.gov}

\and

\author{John F. Hawley}
\affil{Astronomy Department\\
University of Virginia\\ 
P.O. Box 400325\\
Charlottesville, VA 22904-4325}
\email{jh8h@virginia.edu}

\begin{abstract}
Recent general relativistic magneto-hydrodynamic (MHD) simulations of accretion onto black
holes have shown that, contrary to the basic assumptions of the Novikov-Thorne model,
there can be substantial magnetic stress throughout the plunging region.  Additional
dissipation and radiation can therefore be expected.  We use data from a particularly
well-resolved simulation of accretion onto a non-spinning black hole to compute
both the radiative efficiency of such a flow and its spectrum if all emitted light
is radiated with a thermal spectrum whose temperature matches the local effective
temperature.  This disk is geometrically thin enough ($H/r \simeq 0.06$) that little
heat is retained in the flow.  In terms of light reaching infinity (i.e., after
allowance for all relativistic effects and for photon capture by the
black hole), we find that the
radiative efficiency is at least $\simeq 6-10\%$ greater than predicted by the
Novikov-Thorne model (complete radiation of all heat might yield another $\simeq 6\%$).
We also find that the spectrum more closely resembles the Novikov-Thorne prediction for
$a/M \simeq 0.2$--0.3 than for the correct value, $a/M=0$.  As a result, if the
spin of a non-spinning black hole is inferred by model-fitting to a Novikov-Thorne
model with known black hole mass, distance, and inclination, the inferred $a/M$ is too large
by $\simeq 0.2$--0.3.

\end{abstract}

\keywords{accretion, accretion disks --- black hole physics --- MHD --- radiative transfer --- X-rays: binaries}

\section{Introduction}

    When astrophysicists consider accretion onto black holes (BHs), they are
primarily concerned with the associated radiative output because that is what
we observe.  This output can be characterized in summary terms by the radiative
efficiency, the energy in photons generated per unit rest-mass accreted.  It can
also be characterized at a more detailed level by its spectral shape.

The radiative efficiency is valuable because it directly links the measured
luminosity to the key physical parameter governing accretion dynamics,
the accretion rate.  It serves in the same way to translate
the integrated energy of BH light production (dominated by
AGN output) to the mass density of BHs in the Universe \citep{soltan82}.
In fact, it is the expected high radiative efficiency ($\sim O(0.1)$) of
BH accretion that was the original linchpin of the argument that
only BH accretion could explain the luminosities of quasars: at
the much lower efficiency of stellar nucleosynthesis, the mass budget
would be unsupportable.  Spectral shapes can provide much more detailed and
specific diagnostics with which to probe the dynamics of accretion, but creating
that linkage demands much more in the way of microphysics.

     Given its centrality, it is not surprising that the first effort to
calculate the radiative efficiency of BH accretion came very early
in the development of the subject \citep{1973blho.conf..343N,1974ApJ...191..499P}.
The approach of the Novikov-Thorne (NT) model, as it has come to be called,
rested on: two indubitable principles, conservation of energy
and conservation of angular momentum; two symmetry assumptions,
that the accretion flow was time-steady and axisymmetric; and a plausible
physical approximation, that all dissipated heat was radiated away immediately.
It neglected the loss of some photons to capture by the BH, but that
omission was quickly repaired \citep{T74}.  However, an additional boundary
condition was also required in order to close the
system of equations, and that could only be guessed heuristically.  This
boundary condition can be phrased in terms of either the net accreted angular
momentum per unit accreted rest-mass or the $r$-$\phi$ component of the stress
tensor at the innermost stable circular orbit (ISCO).  The choice
made in the NT model was to set the specific accreted
angular momentum to the angular momentum of a test-particle orbit at the
ISCO, which is equivalent to setting the stress to zero at and inside
the ISCO.  With that choice, the NT model predicts a radiative efficiency
that is exactly the binding energy of a test-particle orbit at the ISCO
because the fluid is assumed to be perfectly cold at all times and its
trajectory from the ISCO to the horizon is in exact free-fall.
Since the early 1970s, questions have been raised about both the
physical approximation and the boundary condition.

     One objection to the ``prompt radiation" assumption is that in practice
any accreting fluid must carry at least some heat.  However, if the dissipated
heat is radiated thermally, the resulting equilibrium specific enthalpy is
only slightly greater than unity \citep{pringlerees72,SS73}.  In that case,
the advected heat does not
materially reduce the radiative efficiency.  However, there are instances
in which the advected heat can be significant.  When the accretion rate is
very low, the radiation rate may be so low that the cooling time
becomes long compared to the inflow time \citep{i77,rbbp82,ny94}.  Assuming
that electrons receive heat only by Coulomb scattering with hotter ions,
\cite{fragile09} calculated the radiation rate in such a flow, explicitly
demonstrating how much heat could remain in the plasma.  Working with a
more complete treatment of the ion and electron distribution functions,
\cite{sharma07} showed that the MHD turbulence essential to accretion
automatically transfers a significant amount of heat to the electrons,
but the ions may nonetheless carry enough heat inward to depress
the radiative efficiency.  In the opposite extreme of very high accretion
rate, the disk's optical depth may be so great that photons cannot diffuse
out of the disk in an inflow time \citep{B79,A88}.  Just how much heat can
be retained in this case has not been studied in as great detail as for
the low accretion rate case because to do so requires both knowledge of
the dissipation profile within the accreting matter and solution of the
time-dependent radiation transfer equation in conjunction with a solution
of the dynamical equations.

     The ``zero stress at the ISCO" assumption has also been criticized
\citep{K99,G99}.  The nature of the critique here is not whether the
assumption's validity depends on the value of a parameter, but whether
there is any reason why the onset of orbital instability should suppress
the mechanism creating the stress.  At the time the NT model was invented,
there was no understanding of the stress's physical character;
reasoning about it was therefore necessarily phenomenological and heuristic.
Nonetheless, from very early on \citep{T74}, concerns were voiced
that the heuristic reasoning leading to the
zero-stress assumption would break down if magnetic effects were important.
Since the early 1990s, we have come to understand that the
dominant source of stress in accretion disks is in fact MHD turbulence stirred
by the magneto-rotational instability \citep{BH91,BH98}, so magnetic
forces are central to the issue.
In the last ten years, a large number of numerical simulations, many of
them utilizing a 3-d description of MHD in full general relativity, have
been used to explore the question of to what degree these magnetic stresses
continue across the ISCO
\citep{2003ApJ...599.1238D,GSM04,KHH05,shafee08,NKH10,penna10}.
To the degree that they do, the associated dissipation would likely raise
the radiative efficiency.  In fact, \cite{BHK08} attempted to estimate
the additional radiation (and its effect on the observed spectrum) directly
from their simulation data on the stresses.

     Although stress is intimately connected to dissipation, they need not
take place precisely in the same location.  Consequently, for the purpose
of calculating the radiation rate, it is preferable to treat the emissivity
directly.  With contemporary codes, which are not yet capable of solving the
radiation transfer problem simultaneously with the dynamics in global simulations,
radiation can be treated only when the material is assumed to be optically thin.
This assumption is physically justified when the accretion rate, and therefore
the gas density, is very small \citep{fragile09}; it is merely an {\it ad hoc}
device when the gas density is great enough that prompt and (nearly) complete
radiation of the heat can be expected.  Nonetheless, provided the cooling
time is shorter than the inflow time, a toy-model optically-thin cooling
function may still provide a good measure of the energy available for
radiation.  In a previous study, we implemented just such a device and 
applied it to the case of a spinning BH ($a/M=0.9$) and a moderately
thick (aspect ratio $H/r \simeq 0.13$) disk \citep{NKH09}, finding that,
after allowance for photon capture by the BH, the radiative
efficiency was about $6\%$ greater than the NT radiative efficiency, but
if the flow had radiated all its dissipated heat promptly, the fractional
increase may have been as large as $\simeq 20\%$.

    Some \citep{shafee08,penna10} have argued that thicker disks might
increase the stress (and dissipation) level in the near-ISCO region;
if so, most astrophysical accretion disks would show little in the way of such effects.
In \cite{NKH10}, we showed that when all other variables are held
constant and care is taken to simulate with adequate resolution (see \cite{hgk11}
for a detailed discussion of what ``adequate resolution" means in this context),
the near-ISCO magnetic stress levels are essentially {\it independent} of $H/r$
in Schwarzschild spacetimes, at least for the particular magnetic topology
studied.

     In this paper, we calculate both the radiative efficiency of accretion
onto a non-spinning black hole and the shape of the spectrum in the event
that essentially all of the emitted power is thermal.  Because our simulations
yield detailed data on the time- and spatially-dependent emissivity in the disk,
we can easily translate this data into a local fluid-frame effective temperature.
In addition to its intrinsic interest, the shape of this spectrum has particular
significance because there have been extensive efforts to use the thermal continuum
of Galactic black hole binaries to infer BH spin (e.g.,
\cite{Gou09,Gou10,Steiner10}).  We will
explore how one of these simulations, with its explicit calculation of torque and
dissipation, compare to the traditional Novikov-Thorne predictions often used
in these efforts to measure BH spin.

\section{Calculational Method}\label{sec:calcmethod}

\subsection{Simulation data}

     The data we analyze for this paper are drawn from the geometrically
thinnest of the three high-resolution simulations reported in \cite{NKH10}, the
simulation they designate ThinHR.  {\tt HARM3D}, the code used
in this work, is an intrinsically conservative 3-d MHD code in full general
relativity.  Because it uses a coordinate system based on Kerr-Schild, it is
able to place the inner boundary inside the black hole's event horizon, thus
obviating the need for a guessed inner boundary condition on angular momentum
flow through the disk.  The stress-energy conservation equation is modified to
include an optically-thin cooling function; that is, we write
$\nabla_\nu T^\nu_\mu = - {\cal L}u_\mu$, where $T^\nu_\mu$ is the stress-energy
tensor, $u_\mu$ is the specific 4-momentum, and ${\cal L}$ is
non-zero only when the local temperature is greater than a target temperature
$T_*$.  This target temperature is chosen in advance to keep the disk's aspect
ratio $H/r$ close to a single pre-set value at all radii.  In dimensionless
code units, it is
$T_* \equiv (\pi/2) (R_z/r)(H/r)^2$, where $R_z$ describes the correction to
the vertical gravity due to relativistic effects \citep{NKH10}.

     We took special pains to ensure the numerical quality of the ThinHR simulation.
Every $20M$ in time (we set $G=c=1$, so time has units of
$(M/M_\odot)\cdot 4.9\times 10^{-6}$ s), we checked that throughout the disk body
there were at least $\simeq 6$ cells across the fastest-growing wavelength of
the MRI; i.e., $2\pi v_{Az}/\Omega > 6$, where $v_{Az}$ is the local Alfv\'en
speed counting only the vertical component of the magnetic field and $\Omega$
is the local orbital frequency.  The mass- and time-weighted value of this quantity
was $\simeq 25$, well in excess of this minimum.  As discussed in \cite{hgk11},
by this and several other measures, ThinHR is the best-resolved
global accretion simulation in the literature, and the only one that even
begins to approach convergence.  As \cite{hgk11} also showed, simulations with
more complex initial magnetic field geometry than the nested dipolar loops used in
ThinHR are much more difficult to resolve well.

By examining the time-dependence of the mass interior to several fiducial radii,
\cite{NKH10} determined that the final $5000M$ of the ThinHR
simulation met the
relevant criteria for inflow equilibrium in the inner disk.  We have now also studied the
emissivity as a function of time in this simulation and confirmed that that
period is likewise a statistically stationary state with
regard to photon radiation.  We use that period for time-averages in this paper.

\begin{figure}
\includegraphics[angle=90,width=0.8\textwidth]{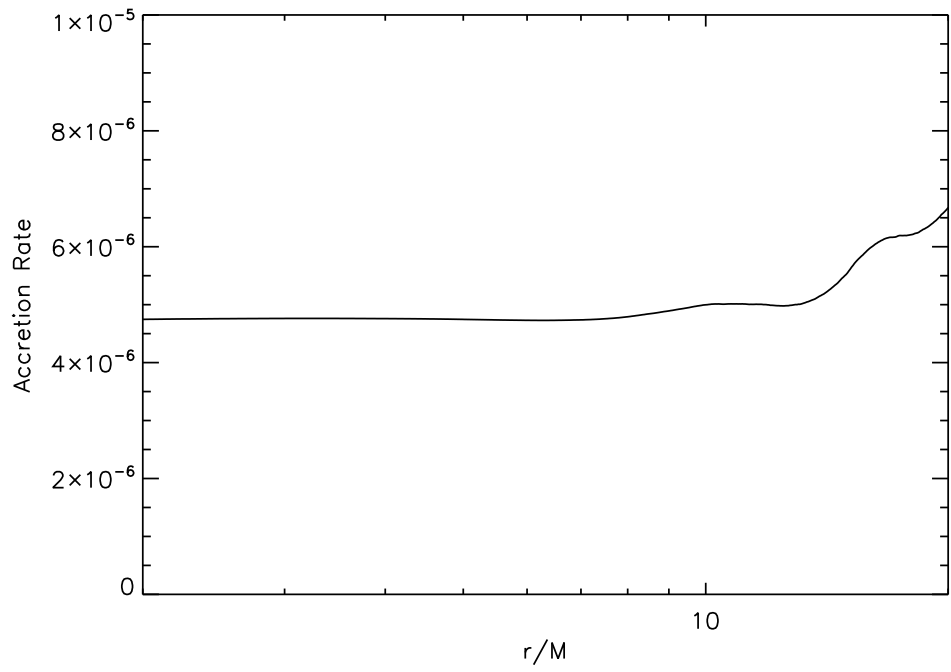}
\caption{Accretion rate as a function of radius averaged over the time of
inflow equilibrium in units of fraction of the initial mass per unit time.
\label{fig:accrate}}
\end{figure}

      In studying simulations intended to represent statistically steady
accretion, it is important to recognize that when there is only a finite
amount of mass on the grid, some of it must move out in order to absorb
the angular momentum removed from accreted material.  Consequently, the
radial range over which the disk can be said to be in inflow equilibrium
is limited.  For the simulation under consideration here, that range
was typically $r \lesssim 20M$ (see Fig.~\ref{fig:accrate}).  The time-averaged
accretion rate is constant to within 5\% for $r \le 14M$, and changes only
by 40\% within $r=20M$.  Because close to half
the total emitted luminosity comes from larger radii, when we evaluate
the total accretion luminosity, it will be necessary to adopt a scheme for
attaching the radiation from larger radii to the part we can compute from
$r \leq 20M$.

      The principal conclusion of \cite{NKH10} was that the time-averaged radial
profile of fluid-frame electromagnetic stress normalized to the time-averaged
accretion rate changed remarkably little as a function of $H/r$.  For the thicker
configurations, there is also a measurable Reynolds stress, but this is always
smaller than the magnetic stress and decreases with disk thickness, so that it is
very small when $H/r=0.06$.
Consequently, there is no reason to believe that the accretion
rate-normalized radial stress profile should show any significant dependence
on H/r in still thinner disks.

\subsection{Approximate solutions of the transfer problem}\label{sec:transferapprox}

The principal goal of this paper is to use the cooling function data from
ThinHR both to compute the radiative efficiency of accretion in this simulation
and to predict the shape of the spectrum in the event that essentially
all the radiation is emitted by locally thermal processes.  Both calculations
depend on the angular distribution of emitted photons.  Because we do not solve the
complete transfer problem inside the disk, we present here two alternate approximations
that, between them, span the range of possibilities.  One of these assumes that
the trajectories of photons from the disk to infinity (or the black hole's event
horizon) can be computed as if the disk material were completely optically thin.
In the other, we assume that the disk is very optically thick, while its surroundings
are perfectly transparent, so that all photon trajectories begin from a photospheric
surface but are exact geodesics for massless particles from there to
infinity (a small fraction also scatter off other regions of the disk
or are captured by the event horizon).

Both versions make use of the same underlying data, the 3-d maps we recorded,
every $20M$ in time, of ${\cal L}$ and $u_\mu$, but with certain adjustments.
As shown in Figure~\ref{fig:accrate}, the accretion flow is very close to inflow
equilibrium for $r \lesssim 20M$.  For those radii, the time-averaged accretion
rate is nearly independent of $r$.  Near $r \simeq 25M$, it rises to a maximum
$\sim 40\%$ greater than the rate at smaller radii; at still larger radii, it
plummets; at sufficiently large radii, the flow turns outward, as it must in
order to convey the outward-directed angular momentum flux.
Because we wish to make radiation predictions for disks that are, on average,
time-steady, yet a significant part of the total luminosity is emitted from radii $>20M$,
we adopt the following procedure: For $r < 20M$, we adjust the
local emissivity ${\cal L}$ by the factor
$\langle \dot M(r=r_{\rm ISCO})\rangle/\langle \dot M(r) \rangle$, where 
$\langle X \rangle$ denotes the time average of $X$.  The very small departures
from inflow equilibrium in this range of radii mean that this adjustment
makes only a very slight change.  For $r > 20M$,
we use the emissivity as predicted by the NT model for
$\dot M = \langle \dot M(r=r_{\rm ISCO})\rangle$.

Both methods also share the assumption that, even though our cooling function is defined
as if the gas were optically thin, the optical depth is large enough to produce an
emergent spectrum that is close to thermal. To find the effective temperature of that local
thermal spectrum, we define the surface brightness as a function of $r$ and $\phi$
by integrating the fluid-frame emissivity over polar angle:
\begin{equation}
S(r,\phi) = (1/2)\int \, dx^{(\theta)} {\cal L}(r,\theta,\phi),
\end{equation}
where differential distances in the fluid frame are found by projection onto a
local tetrad, i.e., $dx^{(\mu)} = \hat e^{(\mu)}_{\nu} dx^\nu$.  The factor of
2 enters because the disk has both a top and a bottom surface.  This surface
brightness is the closest approximation we can readily make to a vertical
integration through a thin disk.  Because ThinHR has an aspect ratio of
only $\simeq 0.06$, there is very little difference between a polar angle
and a vertical integration.
The effective temperature is then
\begin{equation}\label{eq:teff}
T_{\rm eff}(r,\phi) \equiv \left[S(r,\phi)/\sigma\right]^{1/4},
\end{equation}
where $\sigma$ is the Stefan-Boltzmann constant.

The shape of the integrated spectrum at infinity is uniquely determined by
the simulation data, but the dimensionless code data specify neither the units of photon
energy nor of luminosity.   Attaching physical units to the results
demands choosing two parameters.  A particular choice of central mass $M$
determines the units of length and time; a particular value of the accretion
rate $\dot M$ determines the unit of mass in the fluid.  Between these two,
both the unit of photon energy and the unit of luminosity can be determined.
The characteristic photon energy (or temperature) scale is $\propto (\dot M/M^2)^{1/4}$ as
expected from conventional disk theory.  The integrated emissivity is, of
course, $\propto \dot M$.  Technical details regarding the translation between
code units and physical units can be found in the Appendix.

\subsection{Optically thin method}\label{sec:optthin}

    This approximate solution to the radiation transfer problem is, in
some sense, ``truest" to the physics of the simulation. Even though we assume
a thermal spectrum, we permit each cell to radiate isotropically in its frame,
and the photons then travel without hindrance to infinity.

To compute the bolometric luminosity received at infinity, we use the cooling
function snapshots as input to the general relativistic ray-tracing engine described
in \cite{NKH09}.   We trace rays from an array of observers at infinity at different
polar angles to the simulation volume and then integrate over solid angle in order
to determine the total luminosity.

Photon geodesics are defined by
\begin{equation}
{\partial x^\mu \over \partial \lambda} = N^\mu , \qquad 
{\partial N^\mu \over \partial \lambda} = \Gamma^\mu_{\alpha\beta}
                                                N^{\alpha} N^{\beta}
\end{equation}
and the intensity arriving at infinity comes from integrating the invariant
transfer equation
\begin{equation}
{d {\cal I} \over d\lambda} = {\cal J},
\end{equation}
where ${\cal I} = I_\nu/\nu^3$ is the invariant intensity, ${\cal J}$ is
the invariant emissivity, and $\lambda$ is a scalar affine parameter.

When computing the bolometric flux received at infinity, it is convenient to
suppose that all photons received at infinity have the same frequency $\nu_o$, so that
\begin{equation}
{\cal J} = {{\cal L} \over 4\pi \nu^2}\delta(\nu - \nu_o/g),
\end{equation}
with $g$ the Doppler factor relating the emitted fluid-frame frequency to the
frequency at infinity.  To determine the observed luminosity per solid angle
$dL/d\Omega$, we averaged over the time period during which the simulation was
in inflow equilibrium.

For computing the integrated spectrum received by these distant observers, we need
to define the spectrum of the photons radiated by each cell.  So that all cells with
the same radial and azimuthal coordinates produce a spectrum corresponding to the local
effective temperature, we define a spectral emissivity per unit solid angle and frequency
in the fluid frame
\begin{equation}
{\cal L}_\nu = B_\nu (T_{\rm eff})\frac{{\cal L}}{B},
\end{equation}
where $B_\nu$ is the usual Planck function and $B \equiv \int \, d\nu \, B_\nu$.
With this choice, the shape of the locally-emitted spectrum matches that of a black body at
the local effective temperature while its total power matches the local luminosity.

\subsection{Optically thick method}\label{sec:optthick}

Our second approximate solution is ``truer" to the assumption of a thermal
spectrum.  We determine the code unit value of the local effective temperature
exactly as done for the optically-thin method just described.  However, instead
of defining a volume emissivity, we instead use a {\it surface} flux
$\pi f^{-4} B_\nu (f T_{\rm eff})$ and ray-trace the emission to
infinity from a photosphere.  Consistent with this approximation's optically-thick
assumption, we suppose that the atmosphere is scattering-dominated, so that the
spectrum emerges with a Comptonization hardening factor $f = 1.8$ \citep{Shimura:1995} and the
angular distribution of the intensity follows the scattering-dominated
limb-darkening law of \cite{Chandra:1960}.  In principle, the photosphere could be placed
at the midplane and its velocity could be set at the mass-weighted velocity
for that value of $(r,\phi)$, so that the ray-tracing could be done once and for all.
We could then transform the result into physical units exactly as for the optically-thin
method.  However, we instead specify physical units before the ray-tracing so
that the location of the photosphere can actually be calculated in terms of
the density distribution found in the simulation.\footnote{This method encounters a complication near the ISCO.
As the surface density diminishes, the disk can become optically thin.  In
those regions, we define the photosphere as lying in the midplane.}
 
The photosphere typically occurs at several scale heights
above the midplane (as is also seen in shearing box simulations with far more
detailed thermodynamics; see \cite{HKS06}). Since we define a different
photosphere at each point $(r,\phi)$ in the disk, for each frame of
simulation data, the emission and ray-tracing are truly
three-dimensional and dynamic. Only afterwards do we integrate over
azimuth and time to produce the observed spectra. Just like a real
detector, this dynamic, 3-d ray-tracing allows us to accurately model
the effects of isolated hot-spots and velocity perturbations in the
disk that tend to lead to a harder spectrum.

At intervals of $100M$ from $t=10000M$ to $t=15000M$, we used the 3-d simulation data as
boundary conditions for the
general relativistic radiation transfer code described in
\cite{Schnittman:2009} and \cite{Schnittman:2011}.
With that code, for each point on the photosphere ($r \times \phi
= 360\times 64$)
in each snapshot, we followed $\sim 10^4$ photon packets along
outward-directed rays randomly selected in direction with a probability
distribution uniform over solid angle in the fluid frame.  Most
photons reach infinity (here, $r=10000M$).  A minority are captured by the black hole.
Another minority strike the accretion disk somewhere else, where they are scattered
with a redistribution function following the expression for a scattering-dominated
atmosphere derived by \cite{Chandra:1960}.\footnote{This returning radiation contributes
only slightly to the total flux, but can dominate the polarization at the high
energy end of the spectrum (\cite{Schnittman:2009}.}  To determine the flux directed
in a given
solid angle, we chose 41 bins evenly spaced in $\cos\theta$ and grouped all photons
arriving within a single bin.  When we cite a bolometric luminosity $dL/d\Omega$
in a particular direction, it is computed by integrating in frequency
over $dL_\nu/d\Omega$.

\section{Results}

\subsection{Fluid-frame flux}

Figure~\ref{fig:ffflux} displays the time- and azimuthally-averaged surface brightness
of ThinHR, as measured in the local fluid frame (i.e., the local orbital
frame).  The surface brightness follows the Novikov-Thorne model
prediction at $r \gtrsim 10M$ because at such radii, the stress at the ISCO has
little effect---the specific accreted angular momentum (the
parameter fixed in the NT model by the ISCO stress) is small compared to the local
specific angular momentum.  Near and inside the ISCO, however, the surface
brightness contrasts sharply with the NT model.  It remains at a high level
all the way to the event horizon.

\begin{figure}
\includegraphics[angle=90,width=0.8\textwidth]{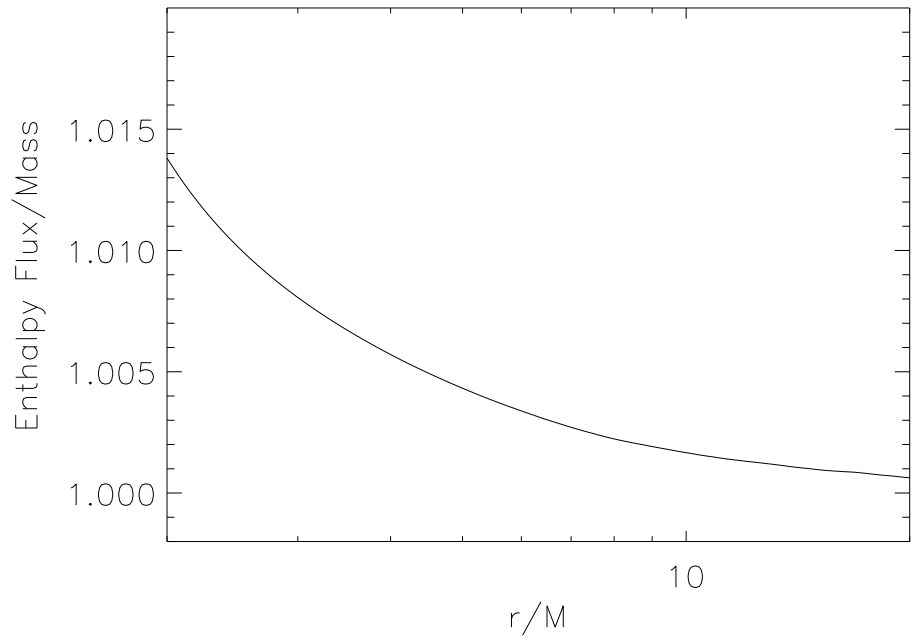}
\caption{Time-averaged enthalpy per unit rest-mass in the accretion flow.
\label{fig:enthalpy}}
\end{figure}

Our results may also be compared to those of \cite{BHK08}.  Scaling dissipation rate
to stress, they predicted fluid-frame dissipation rates that rose steeply through the
plunging region.  One reason our plunging region emissivity is somewhat less
than their estimate is that our cooling function does not lead to complete radiation
of all dissipated heat.  Adiabatic expansion can lower the temperature below the
target temperature despite continued dissipation.  As a result, some heat is kept
in the fluid all the way to the horizon.  The amount of heat
retained is illustrated in Figure~\ref{fig:enthalpy}, which shows the time-averaged
enthalpy per unit rest-mass in the accretion flow
\begin{equation}
\langle h \rangle(r) = \left\langle \int \, d\phi \, d\theta \sqrt{-g} \rho h u^r /
        \int \, d\phi \, d\theta \sqrt{-g} \rho u^r \right\rangle .
\end{equation}
At the ISCO, $\langle h \rangle \simeq 1.0034$; at the horizon it rises to 1.014.
For a sense of scale, the excess enthalpy at the ISCO is about 6\% of the radiated
energy.  At $r \simeq 3.5M$, $\langle h \rangle \simeq 1.007$, but the photon capture
probability that deep in the potential is $\simeq 50\%$.  Thus, the unradiated heat
available to reach infinity might be $\simeq 0.0035$ in rest-mass units.

\subsection{Luminosity at infinity}

\begin{figure}
\epsscale{1}
\plotone{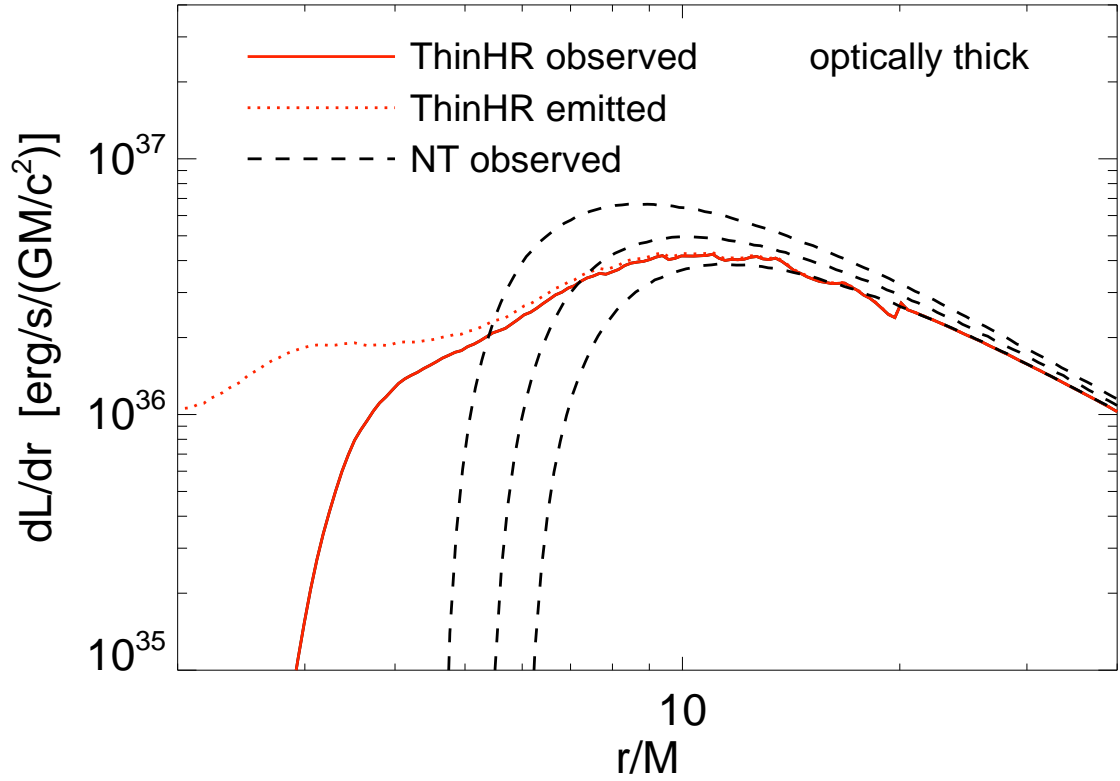}
\caption{Solid-angle integrated luminosity per unit radial coordinate $dL/dr$ for
ThinHR (red curves) contrasted with NT predictions (black dashed curves)
for the same time-averaged accretion rate.  The solid red curve shows the luminosity
reaching infinity; the dotted red curve shows what that luminosity would have been
if no photons were captured by the black hole.  The three black dashed curves
represent spins $a/M = 0$, 0.2, and 0.4 (bottom to top).
\label{fig:dldr}}
\end{figure}

   We used both ray-tracing techniques to translate
the fluid-frame emissivity into luminosity received at infinity as a function of
the radius from which it was emitted (Fig.~\ref{fig:dldr} shows the optically-thick
version; the optically-thin differs only very slightly).  Because the
code unit of gas density in the simulation is arbitrary and all distances and
times are in units of $M$, the simulation data could be used to predict the
luminosity from an accretion flow onto a BH of any mass and any
accretion rate.  All that is necessary is to scale the luminosity appropriately.
For purposes of illustration, we have chosen $M=10M_{\odot}$ and $\dot m = 0.1$,
where $\dot m$ is the accretion rate in Eddington units, because these numbers
are representative of the thermal state in Galactic BH binaries.
Throughout the remainder of this paper, all results presented in cgs units
assume these parameters.

\begin{figure}
\epsscale{1}
\plotone{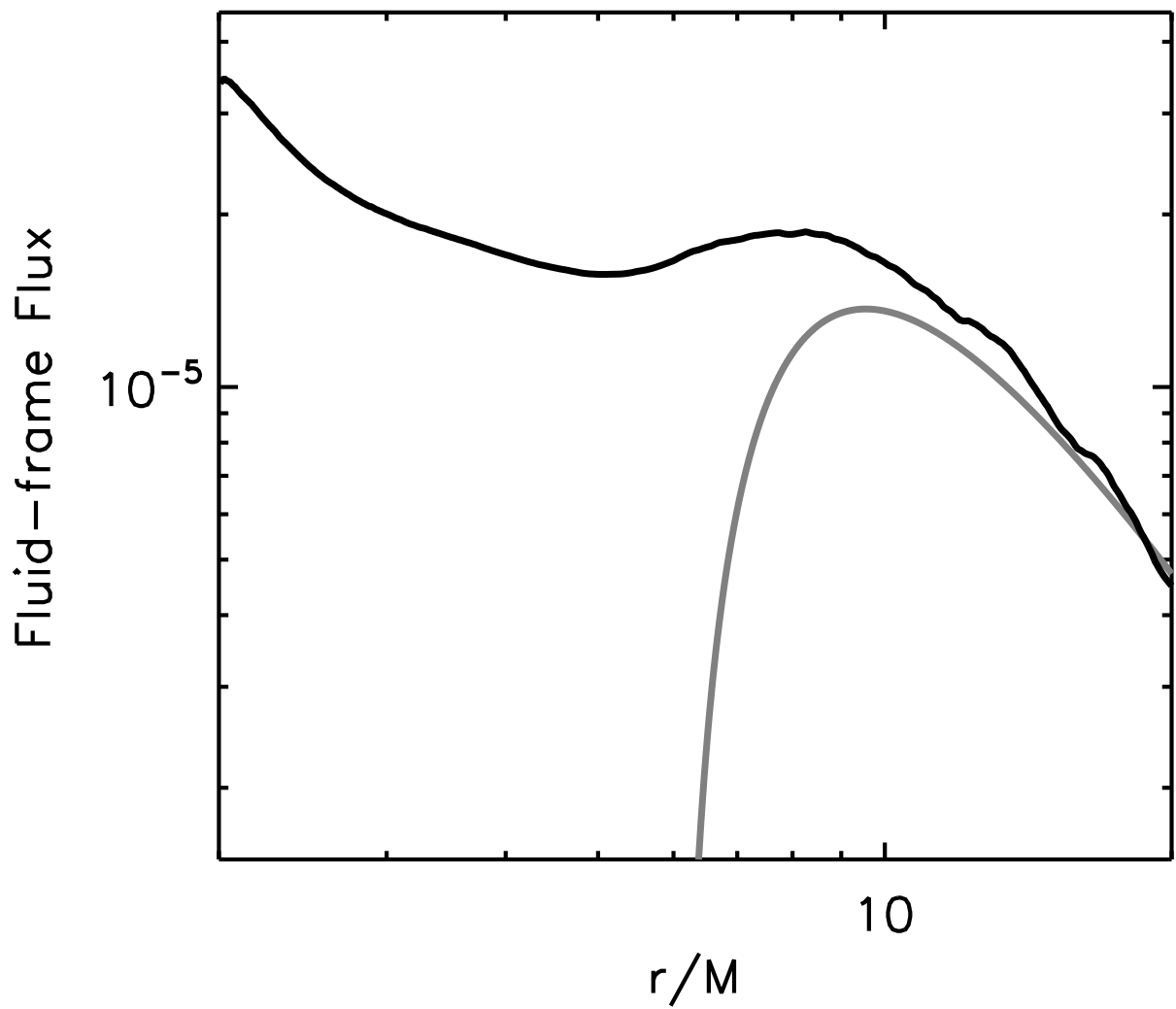}
\caption{Fluid frame flux, averaged over azimuthal angle and over the time of inflow
equilibrium for ThinHR.  The NT model prediction for $\dot{M}=1$ is shown as a light gray curve. 
ThinHR's flux includes a factor $1/\langle \dot{M}(r) \rangle$ to compensate for its 
slight deviation from perfect inflow equilibrium. 
\label{fig:ffflux}}
\end{figure}

The shape of the luminosity profile is, of course, entirely independent of these
specific choices.  To ease comparison between Figure~\ref{fig:dldr} and
Figure~\ref{fig:ffflux}, we present it as
$dL/dr$, the solid angle-integrated luminosity produced
per unit radial coordinate as a function of $r$.  Not surprisingly, in
the disk body $(r \gtrsim 10M)$,
where relativistic effects are smaller, there is little difference
between the simulation prediction and the NT prediction; the small differences
appearing are likely
due to the fact that our averaging time is not
sufficient to smooth away all fluctuations.

     From $r \simeq 10M$ inward, the $dL/dr$ profile mimics the fluid-frame surface
brightness profile and stronger contrasts with NT predictions develop:
by $r \simeq 7M$, the observed luminosity per unit radius matches the $a/M = 0.2$
NT prediction and exceeds the zero-spin NT prediction by a factor of 2.
However, deeper in the plunging region photons escaping to infinity suffer
increasingly larger gravitational redshifting and an increasing fraction of emitted
photons is captured by the black hole.  Consequently,
$dL/dr$ is reduced below the surface brightness curves of Figure~\ref{fig:ffflux}
by larger and larger factors as $r$ becomes smaller and smaller.  Inside $r \simeq
3.5M$, the majority of the light emitted from the accretion flow is captured,
and inside $r \simeq 3M$, hardly any reaches infinity.

     Nonetheless, despite the relativistic losses, the range of radii
responsible for generating significant luminosity at infinity extends to significantly
smaller radii than predicted by the NT model.  $dL/dr$ falls only a factor of 3
from its peak (near $r \simeq 10M$) to $r=4M$.   This behavior is in sharp contrast
to the NT model, for which even at $r=7M$, still outside the ISCO, $dL/dr$ is already
reduced by more than a factor of 3 relative to its value at $r=10M$
and is dropping fast with decreasing radius.

\subsection{Integrated radiative efficiency}

The classical Novikov-Thorne prediction for the radiative efficiency of a
perfectly-radiating disk around a Schwarzschild BH is
$1 - \sqrt{8}/3 \simeq 0.0572$, the binding energy at the ISCO.  Although few
photons radiated this far out are captured by the BH, this number
must still be corrected for such effects.  If the photons are radiated isotropically
in the local fluid frame (i.e., our ``optically thin" approximation), the actual
efficiency of photons reaching infinity falls to 0.0553; if they are radiated
with an angular distribution corresponding to an optically thick disk having
the limb-darkening of a scattering-dominated atmosphere, it is a bit larger,
0.0570 (the lower efficiency of an optically thin NT disk is due
to the larger photon capture rate, with more photons initially emitted
in the plane of the disk).

By contrast, we find that the radiative efficiency predicted by a simulation
incorporating MHD dynamics is somewhat larger: 0.0608 for the optically thin
model, 0.0606 for the optically thick (now the optically thin model
leads to {\it higher} efficiency, since there is enhanced emission near the
ISCO, where velocities are higher and more energy gets beamed in the
forward direction towards high-inclination observers).
These amount to an increase in efficiency of $\simeq 10\%$
for the optically thin case or $\simeq 6\%$ with the angular distribution of
an optically thick disk.  As we remarked above, these numbers might in principle
be increased by another $\simeq 6\%$ if a larger fraction of the
fluid's heat were radiated.

\begin{figure}
\epsscale{1}
\plotone{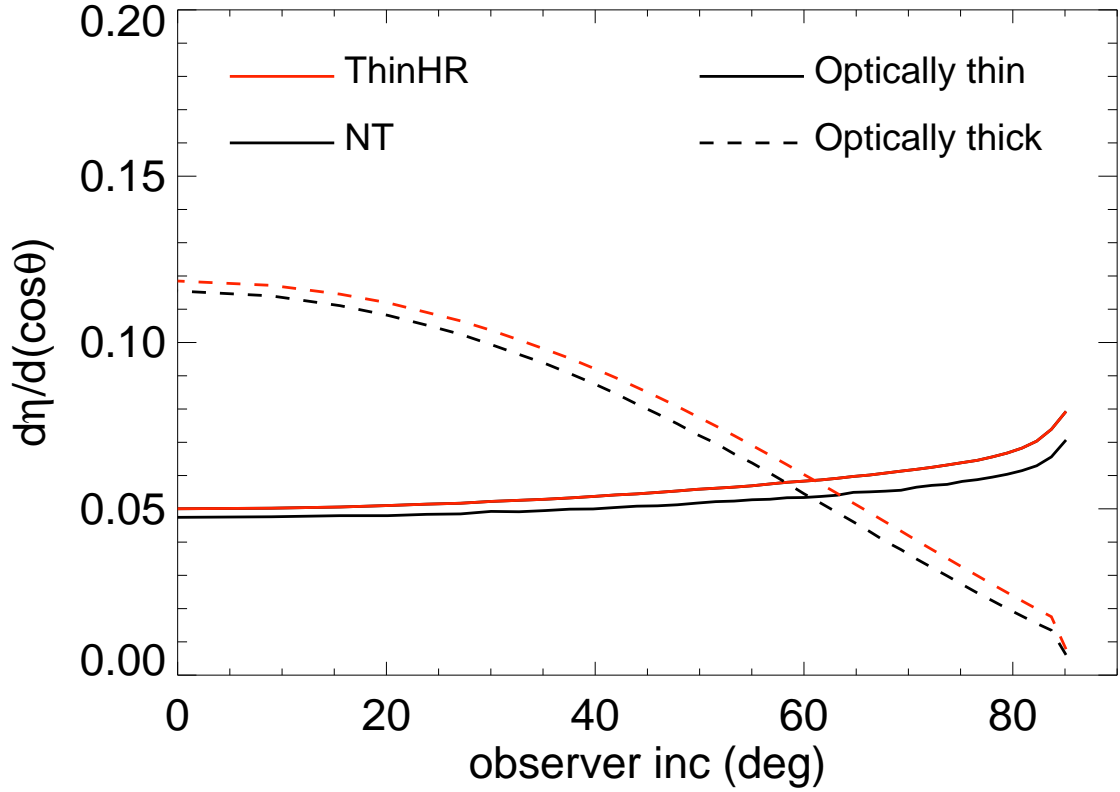}
\caption{Luminosity per solid angle per accreted rest-mass, i.e., radiative efficiency
at infinity, as a function of viewing angle.  Simulation data
is shown in red, the NT model (for equal accretion rate) in black.  Solid
curves show the angular dependence
if the light is emitted isotropically in the fluid frame (i.e., assuming the
gas is optically thin), dashed curves show the angular dependence if the light
is radiated from a geometrically thin, optically thick surface.
\label{fig:dldOmega}}
\end{figure}

Because of geometric projection, limb-darkening in the disk atmosphere,
and relativistic beaming effects, the perceived efficiency varies with
viewing angle.  Our two ray-tracing methods differ in their assumptions
about the disk's opacity, and therefore differ in their predictions for
this angular dependence.  This contrast is illustrated in Figure~\ref{fig:dldOmega}
in radiative efficiency units.  The radiation is nearly isotropic in optically thin
conditions; the only angular variation is that $d\eta/d\Omega$
rises by $\sim 50\%$ at high inclination angles as a result of Doppler boosting
and beaming by the innermost orbiting matter.  This effect is slightly stronger
for ThinHR than for the NT model because its emissivity extends to
smaller radii where velocities are greater. 
On the other hand, the optically thick assumption leads to an angular dependence
dominated by the $\cos\theta$ area projection, limb darkening, and the
finite thickness of the disk, which gives $\eta(\theta \gtrsim
85^\circ)=0$ due to self-eclipsing.  In the optically thick case, the perceived
efficiency can vary from a maximum $\simeq 0.12$ (face-on) to essentially nil (edge-on).

\subsection{Predicted Thermal Spectrum}

\begin{figure}
\vbox{\includegraphics[width=0.7\textwidth]{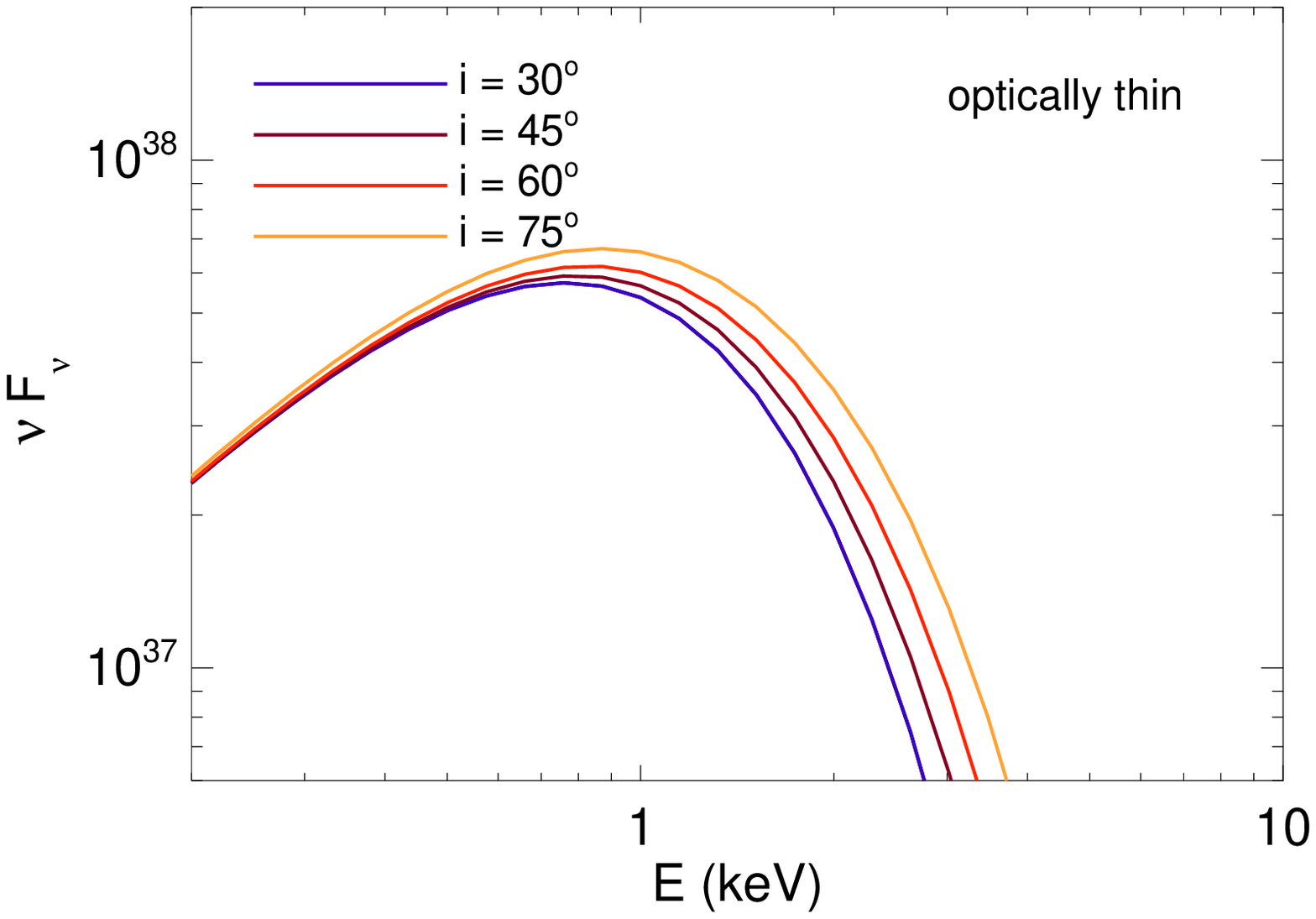}
\includegraphics[width=0.7\textwidth]{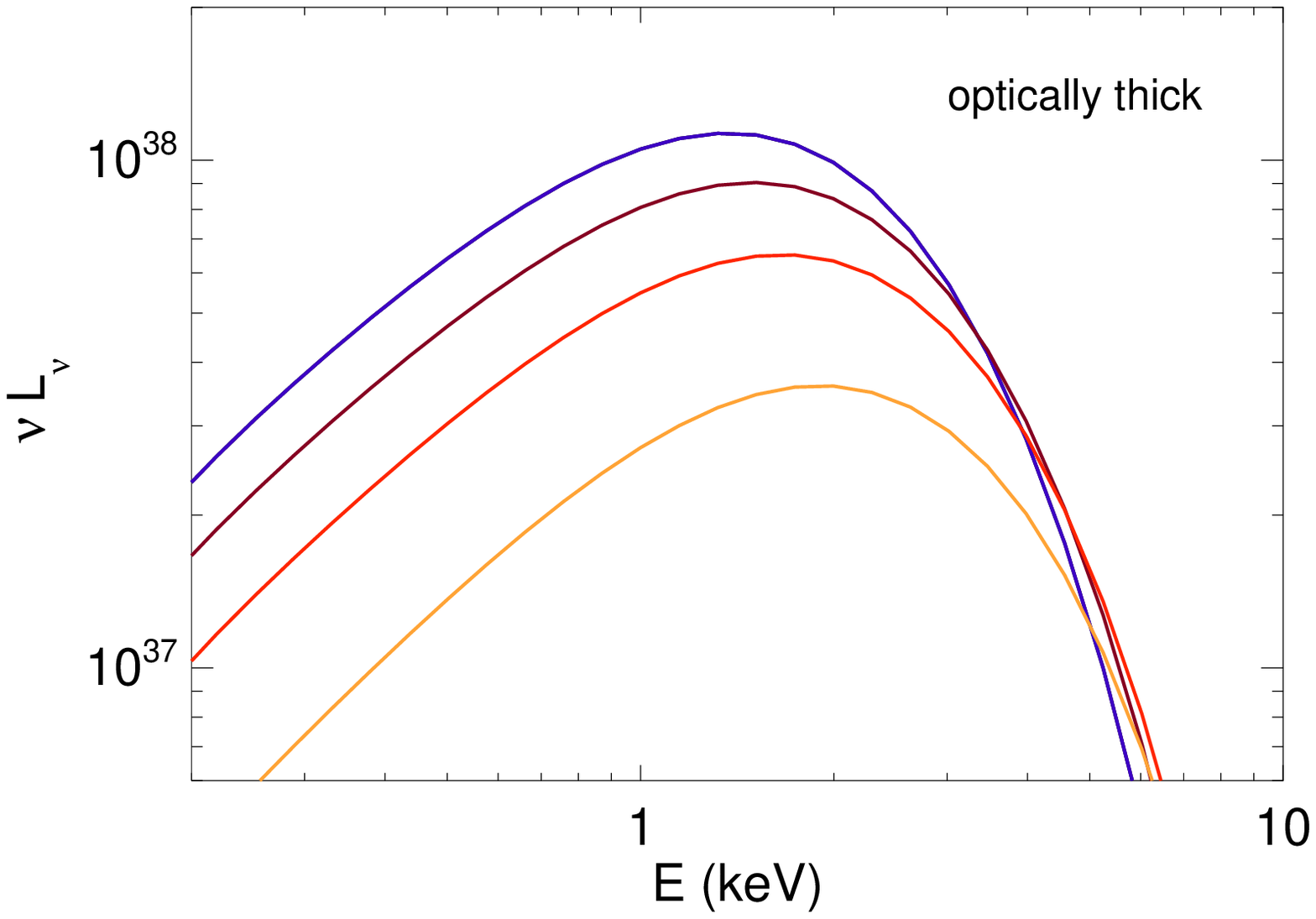}}
\caption{Spectra predicted by our model at several different viewing angles.
(Top) In the optically thin model.  (Bottom) In the optically thick model.
\label{fig:thermspec_ang}}
\end{figure}

Using the methods described in \S~\ref{sec:calcmethod}, we have computed the
spectrum emitted if all the radiation is emitted thermally.    In the two panels of
Figure~\ref{fig:thermspec_ang}, we show how both the optically thin and optically
thick approximations would appear at a selection of viewing angles.  Because
optically thin radiation is close to isotropic, any dependence on viewing angle
is limited.  Only at the highest energies, where the photons predominantly
come from the most relativistic portions of the flow, is there any angle dependence;
above 1~keV (for these parameters), the flux increases with higher inclination.
Optically thick radiation shows a strongly contrasting picture.  In this case, face-on
views yield considerably greater flux, particularly at lower photon energies.

The optically thin and optically thick assumptions also differ in their prediction
for where the $\nu F_\nu$ spectrum peaks.  In the former case, it is (again,
for these parameters) between 600~eV and 1~keV; in the latter case,
due to the spectral hardening from atmosphere scattering, the peak
comes at higher energies, between 1.5 and 2~keV, with higher inclination angles
generating noticeably harder spectra.

\begin{figure}
\epsscale{1}
\plotone{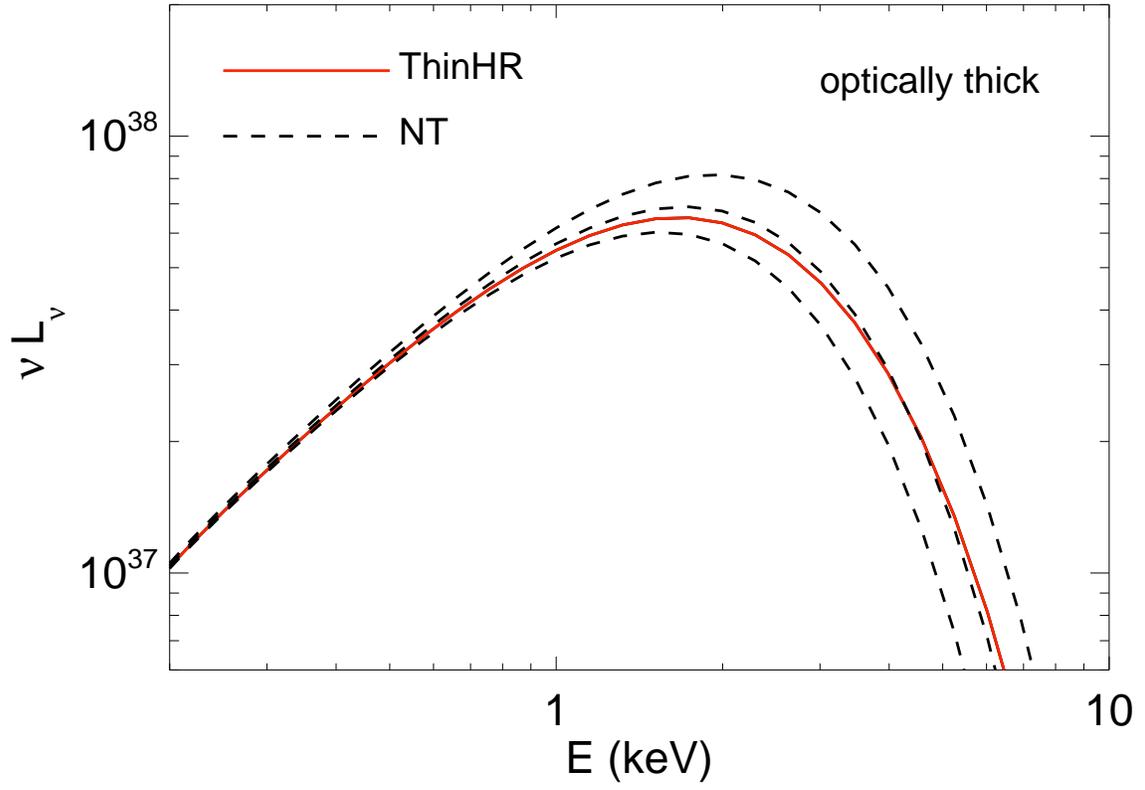}
\caption{Spectrum as seen at $60^\circ$ inclination according to our optically-thick
model (solid red curve) and according to three (optically thick)
Novikov-Thorne models with the same
accretion rate as the simulation (dashed curves) for spins $a/M = 0, 0.2, 0.4$
(bottom to top).
\label{fig:thermspec_nt}}
\end{figure}

In Figure~\ref{fig:thermspec_nt}, we contrast the spectrum seen at an inclination
of $60^\circ$ in the optically-thick model to the spectrum predicted at that
inclination for several optically-thick Novikov-Thorne models of varying
spin.  Not surprisingly,
at photon energies well below the peak, there is very little difference, either
due to the additional physics of our simulation or to the effects of black
hole rotation.   Near the peak the curves begin to diverge, with the spectrum
predicted from the simulation data systematically brighter at higher photon
energies than the zero-spin NT model would predict.  Where $\nu L_\nu$ is the
greatest, the simulation predicts a luminosity greater than NT by $\simeq 20\%$; at
photon energies three times greater, the discrepancy is a factor of 2.

Roughly speaking, the inward extension of high surface brightness due to continued
dissipation near the ISCO resembles the inward extension of high surface brightness
in the NT model when the spin increases.  In this case, the nearest match is to
$a/M \simeq 0.2$.  The degree to which a higher-spin NT model can mimic MHD turbulent
dissipation will be examined more quantitatively in the next section.

\section{Systematic Errors in Inferring Spin from Continuum Spectral Fitting}

Given the difference between the surface brightness profile predicted by our physical
simulation and that of the NT analytic model with its guessed inner boundary
condition, one might well expect that forcing a fit to the NT prescription might
result in significant systematic error in the inferred spin.  In this section, we
evaluate the character of that error.

If the accretion rate were known, Figure~\ref{fig:thermspec_nt} already shows that
a forced NT fit would tend to suggest a spin somewhat greater than the actual one.
However, that is never the case; the accretion rate is also a free parameter.
Moreover, it is hardly the only one: in most cases there are uncertainties about
the inclination,
and sometimes significant uncertainty about the distance and BH mass as well.
The sense and magnitude of the systematic error can also be affected by the process
of simultaneously fitting for these parameters.

To quantify these effects without restricting ourselves to the properties of any
particular instrument or measurement, we begin by defining a quality-of-fit parameter
modeled after $\chi^2$:
\begin{equation}
X^2 = \sum_i \frac{(F_i^{NT} - F_i^{sim})^2}{\sigma_i^2},
\end{equation}
where $F_i^{NT}$ and $F_i^{sim}$ are the spectral predictions by the NT model and
the simulation in $F_\nu$ units, and $\sigma_i$ is the measurement ``error"
in that bin.  The number of photons per bin is $\propto F_i \Delta\nu/\nu$;
if only Poisson errors are relevant, $\sigma_i^2 \propto F_i$ provided that
$\Delta \nu/\nu$ is constant.  In the fits we will show here, the range of
energies considered was 0.2--10~keV.  We have also experimented with restricting
that range to 2--10~keV in order to more closely resemble existing instruments
such as {\it RXTE}; although the ability to distinguish different models does suffer
somewhat, none of our qualitative conclusions is altered.

Consider first the ability to fit simultaneously for all possible unknown parameters:
BH mass, distance, inclination, spin, and accretion rate.  As in our
previous illustrative examples, we choose a case in which the actual mass is
$10M_{\odot}$, the accretion rate is 0.1 in Eddington
units, and the spin is the spin of the ThinHR simulation, i.e., $a/M =
0$. We consider three different target inclinations: $15^\circ$, $45^\circ$, and
$75^\circ$. Given
the freedom to adjust the BH mass, distance,
and accretion rate independently, we find that it is possible to find excellent fits
across virtually the entire spin--inclination angle plane.  That is, with this
many free parameters, one can neither distinguish the NT surface brightness profile
from the simulation prediction nor, assuming the NT model, come close to determining
any of the parameters.

\begin{figure}
\begin{center}
\epsscale{1}
\begin{tabular}{ll}
\includegraphics[width=0.5\textwidth]{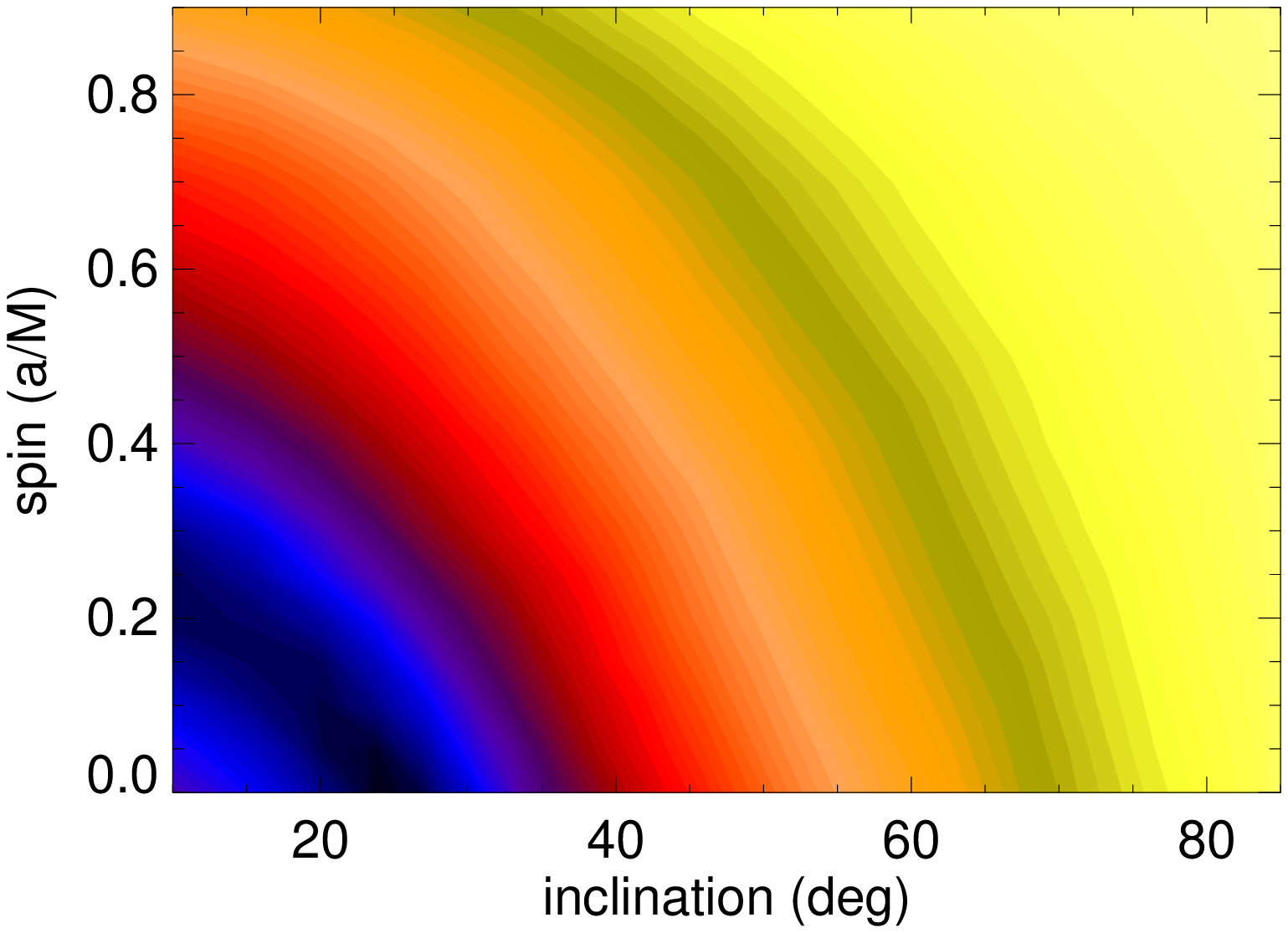}\\
\includegraphics[width=0.5\textwidth]{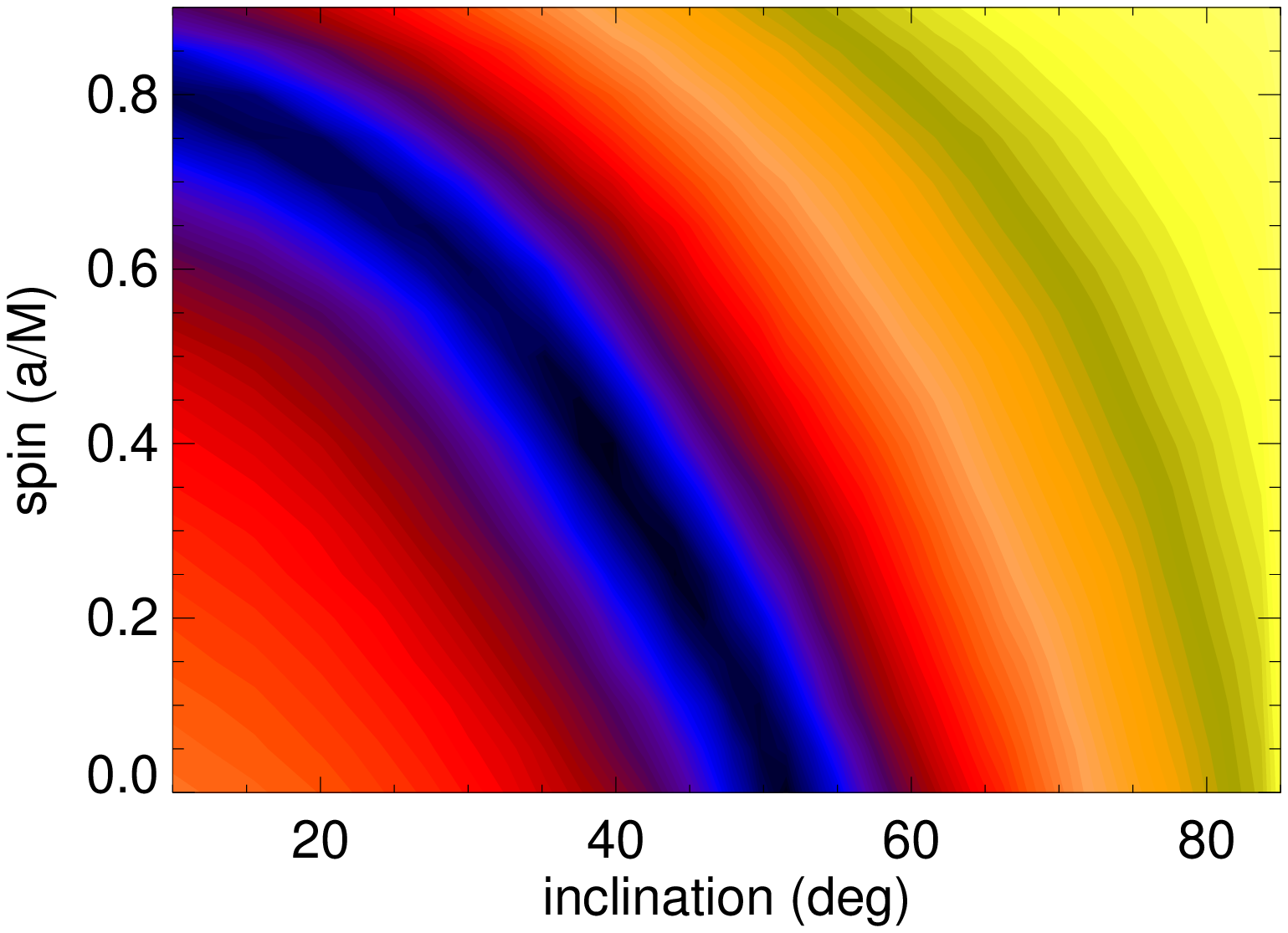} & \includegraphics[width=0.4\textwidth]{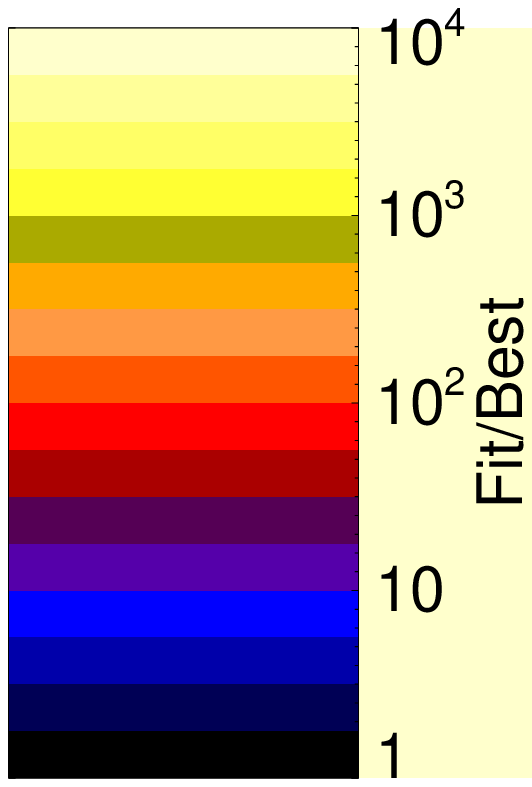}\\
\includegraphics[width=0.5\textwidth]{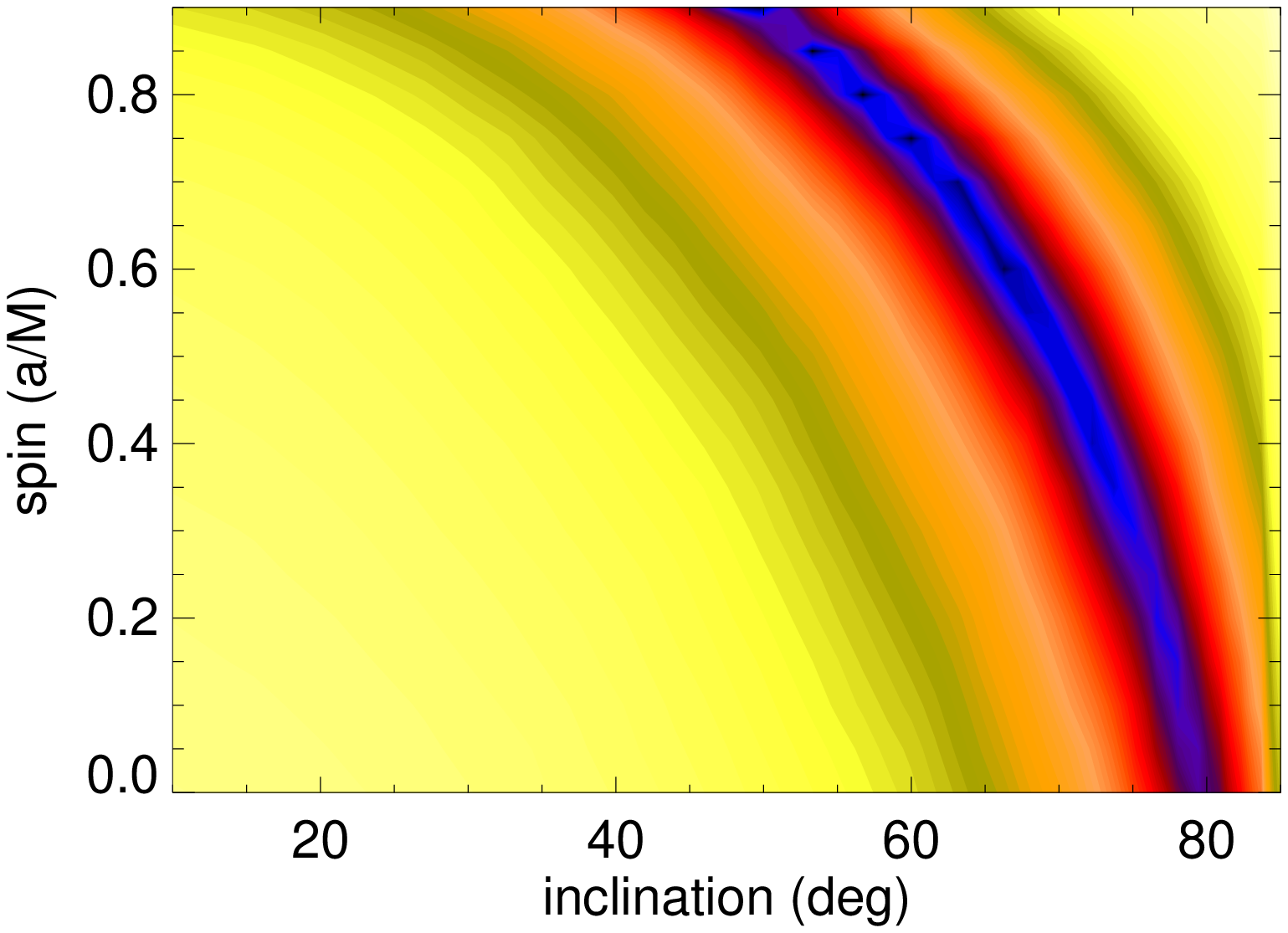}
\end{tabular}
\caption{Fit quality ($X^2$) as a function of BH spin and inclination angle
when the mass and distance are known, but the accretion rate may be
freely adjusted. From top-to-bottom, the target inclinations are
$15^\circ$, $45^\circ$, and $75^\circ$.
\label{fig:pspace2A}}
\end{center}
\end{figure}

Because there are often decent constraints on the mass and distance
(see, e.g., \citep{Orosz:1997,Miller-Jones:2009,Orosz:2011}), it is also relevant
to consider the case in which they are fixed.  As can be seen in
Figure~\ref{fig:pspace2A}, if one attempts to fit a NT model to data derived from
the simulation, one can still find good fits for an extremely broad range of
spin, but only for values of the inclination angle
that are tightly linked to the spin.  

Much of this spin--inclination angle ambiguity is entirely independent of the
details of the surface brightness profile.  The higher energy portion of the
spectrum comes largely from the inner disk.  Radiation from smaller radii can be
made brighter either by the enhanced Doppler beaming and boosting of large
inclination angles or by higher spin.  In this respect, the inability
of the continuum fitting
method to distinguish $a/M=0$ from $a/M=0.9$ has little to do with the contrast
between the NT model and the simulation prediction.

Where MHD physics does make a difference is the offset between the spin--inclination
angle track and the true underlying parameters.  The allowed track in the inclination
angle--spin plane does {\it not} include the correct values of these parameters.
If one imposes the true value of the spin, the inclination angle inferred is too
large by $\simeq 5^\circ$--$20^\circ$; if one imposes the true value of the inclination,
the inferred spin is too large by $\simeq 0.2$--0.3, the offset increasing slowly
with inclination.

The orbital inclination of stellar-mass BHs can in favorable cases be constrained quite
accurately by modeling the optical/IR light curves of the companion
star \citep{Orosz:1997,Orosz:2011}. However, the {\it orbital} inclination is not
necessarily the same as the {\it disk} inclination, which is what really determines
the shape of the X-ray spectrum.  Future measurements with X-ray polarimeters may be
able to measure the disk inclination directly \citep{Li:2009,Schnittman:2009}, and
thus better determine the BH spin.

Ours is not the first attempt to quantify potential systematic errors
due to departures from the NT model.  In particular, \cite{Kulkarni11} used the
simulations of \cite{penna10} as the basis for such a study.  Their methods
differed in certain particulars from ours.  For example, even though the dependence
of emissivity on temperature is highly nonlinear, they first averaged dynamical
snapshot data over azimuth and time, and then computed spectra from that averaged
data.  This procedure underestimates the predicted flux due to high-temperature
regions localized in either azimuth or time.  They also
used simulation data only for $r \leq 8M$; at larger radii, their spectra were
computed from the NT model (see Fig.~\ref{fig:dldr} for a standard of comparison).
The \cite{penna10} simulations are also significantly
coarser in resolution than the ThinHR simulation employed here.  Perhaps as a
result of these several contrasts, \cite{Kulkarni11} found smaller differences
between NT model spectra and simulation-based predictions than we do.

\section{Summary}

We have used the highest-quality general relativistic MHD simulation data available in order
to estimate the radiative properties of accretion onto a non-spinning black hole.  Because
magnetic stresses do not disappear inside the ISCO, and because the accreting fluid always
retains at least a small amount of heat, the amount of energy available for radiation
is not exactly equal to the binding energy of a test-particle orbit at the ISCO, the
assumption of the classical Novikov-Thorne model.  Moreover, even if it were, the
radiation reaching infinity must be adjusted for the fraction of photons captured by
the black hole.

By tying angular momentum transport directly to physical stresses, and radiation directly
to heating caused by the accretion dynamics, and then tracing the trajectories of emitted
photons from the flow to distant observers, we are able to arrive at a more physically
complete description of radiation associated with accretion onto a black hole.  Our
principal findings are:

First, for disks around non-rotating black holes, the total emitted power
received by distant observers per unit accreted rest-mass is at least $\simeq 6$--10\%
greater than predicted by the Novikov-Thorne model; if all the heat content of the
accreting gas were radiated, this radiative efficiency might increase by another
$\simeq 6\%$.  Most of this extra light
comes from the region near and immediately outside the ISCO, although the region
of significant emissivity does extend inside that radius.  Thus, the regions of
the disk in which stronger relativistic effects may potentially be observed are
brighter than predicted by NT.

Second, the angular distribution of flux depends strongly on assumptions about radiation
transfer within the disk.  If the disk is geometrically thin and very optically thick,
and especially if its atmosphere is limb-darkened by scattering, it is brightest in
the face-on direction.  On the other hand, to the degree that the disk is optically
thin, its flux is more nearly isotropic, or even enhanced in the edge-on direction
by relativistic boosting and beaming acting on the innermost emission regions.  Because
it is possible that in real disks the most relativistic regions are optically thin
even while the outer disk is optically thick, real disks may be hybrids of
these two limits.  At higher (prograde) spin, relativistic effects may enhance the
flux at high inclination even when the disk is optically thick throughout.

Third, if all the radiation is produced thermally, the summed spectrum is noticeably
harder than the classical prediction.  This is perhaps our most important result,
as it provides the most immediately observable contrast with previous expectations.

The spectral contrast is also of interest because of the expected association of
higher-temperature spectra with black holes of higher spin.  This is the principal
effect on which attempts to measure spin from thermal state spectra are based
(e.g., \cite{Gou09,Gou10,Steiner10} for a number of Galactic stellar-mass BHs,
\cite{Czerny11} for an AGN).  To the extent that it is instead an indicator of
additional radiation from deeper in the potential, these inferences of spin may
suffer from a systematic bias if they assume the radial surface brightness profile
is identical to the NT prediction.

This bias is not a simple one to describe because the surface brightness profile
is not the only unknown when fitting spectral models to measured spectra---in general
the mass accretion rate is never known, and the BH mass, distance, and inclination
are often not well-constrained.  Moreover, there are numerous additional
possible sources of systematic error, most of which are outside the scope of this
paper (e.g., the detailed structure of the disk atmosphere, and therefore the detailed
emergent spectrum and limb-darkening: see \cite{DDB06} for a lengthier discussion of
these issues).  Nonetheless, its general sense is to bias NT-modeling toward
spins larger than the true value; for $a/M = 0$, we have shown that this shift
is typically $\simeq 0.2$--0.3.

It is of obvious interest to extend these studies over a broader range of
BH spins.  Better data on the relation between radiative efficiency and spin will
permit, for example, a reevaluation of the inferred mean
spin of AGN obtained from a comparison between the masses of contemporary supermassive
black holes and the integrated light of AGN \citep{yutremaine02,elvis02,volonteri05,wang09}.

Efforts to infer individual BH spin from thermal spectra will likewise benefit from
closer attention to potential systematic errors arising from many sources (disk
atmosphere models, inclination offsets between the orbit and the disk, $\ldots$) including,
as we have emphasized here, from the use of inappropriate models for the radial
surface brightness.

\acknowledgements{
This work was partially supported by NSF grants AST-0507455 and AST-0908336 (JHK), 
NASA grant NNX09AD14G and NSF grant AST-0908869 (JFH), and AST-1028087 (SCN).
The ThinHR simulation was carried out on the Teragrid Ranger
system at the Texas Advance Computing Center, which is supported in part 
by the National Science Foundation.  Some of the post-process ray-tracing was run on
the Johns Hopkins Homewood High-Performance Computing Center cluster.
}

\bibliography{bib}


\section{Appendix}\label{sec:appendix}

The translation from code units to physical units is not entirely trivial.  To keep
the bookkeeping clear, we label quantities in code units with the subscript `cu' and
quantities in cgs with the subscript 'cgs'.

In both the optically thick and optically thin methods, it is convenient to define a
dimensionless photon energy $y = \epsilon/\epsilon_0$, with
\begin{equation}
\epsilon_0 = k \left[ \frac{\dot M_{cgs} c^6}{2\sigma (GM_{cgs})^2 \dot M_{cu}}\right]^{1/4}.
\end{equation}
Any particular solution can then easily be shifted in photon energy by the appropriate
amount.

In the optically thin method, we also need a units translation for the local spectral
emissivity per unit volume.  This is most easily done through the mixed-units
quantity
\begin{equation}
j_y = {15 \over \pi^4} {c^8 \over (GM_{cgs})^3} {\dot M_{cgs} \over \dot M_{cu}}{\cal J}_y,
\end{equation}
where the dimensionless emissivity ${\cal J}_y$ is given by
\begin{equation}
{\cal J}_y= {{\cal L}_{cu} \over \int \, d\hat\theta_{cu} {\cal L}_{cu}}
      {y^3 \over e^{y/{\cal T}} - 1}
\end{equation}
for
\begin{equation}
{\cal T} = \left(\int \, d\hat\theta_{cu}\, {\cal L}_{cu}\right)^{1/4}.
\end{equation}
The entire solution can then be computed for unit values of $M_{cgs}$ and $\dot M_{cgs}$,
but scaled afterward when actual values are chosen.

\end{document}